\documentclass{aastex631}

\usepackage{natbib}
\usepackage{graphicx}
\usepackage{epstopdf}
\usepackage[caption=false]{subfig}
\usepackage{url}
\usepackage{float}

\shorttitle{Exploring physically-motivated models}
\shortauthors{Poolakkil et al.}

\graphicspath{{./}{figures/}}
\begin{document}
\title{Exploring Physically-Motivated Models to Fit Gamma-Ray Burst Spectra}

%% load definitions
\newcommand{\vdag}{(v)^\dagger}
\newcommand\Notes[1]{\textcolor{red}{#1}}

\newcommand{\tte}{\texttt{TTE}\xspace}
\newcommand{\ctime}{\texttt{CTIME}\xspace}
\newcommand{\cspec}{\texttt{CSPEC}\xspace}
\newcommand{\drm}{\texttt{DRM}\xspace}
\newcommand{\drms}{\texttt{DRM}s\xspace}
\newcommand{\lle}{\texttt{LLE}\xspace}

\newcommand{\Cstat}{$C_{\textrm{stat}}$\xspace}
\newcommand{\PGstat}{$PG_{\textrm{stat}}$\xspace}
\newcommand{\Pstat}{$P_{\textrm{stat}}$\xspace}
\newcommand{\chisq}{$\chi^{2}$\xspace}
\newcommand{\chisqdof}{$\chi^{2}_{\nu}$\xspace}

% GRB
\def\Epk{E$_{\textrm{peak}}$}
\def\Eiso{E$_{\textrm{iso}}$}
\def\Liso{L$_{\textrm{iso}}$}
\def\t90{T$_{\textrm{90}}$}
\def\tvar{$t_{\textrm{var}}$}
\def\t0{$t_{0}$}
\def\nufnu{$\nu F_{\nu}$}
\def\ra#1#2#3{#1$^{^\textrm{h}}$#2$^{^\textrm{m}}$#3$^{^\textrm{s}}$}
\def\dec#1#2#3{#1$^\circ$#2$'$#3$''$}
% GRB units
\def\fluence{\textrm{erg}\cdot\textrm{cm}^{-2}}

%%Authors
%
% Contributing authors (ranked)
%
\author[0000-0002-6269-0452]{S.~Poolakkil}
\affiliation{Department of Space Science, University of Alabama in Huntsville, Huntsville, AL 35899, USA}
\affiliation{Center for Space Plasma and Aeronomic Research, University of Alabama in Huntsville, Huntsville, AL 35899, USA}
\author[0000-0003-1626-7335]{R.~Preece}
\affiliation{Department of Space Science, University of Alabama in Huntsville, Huntsville, AL 35899, USA}
\author[0000-0002-2149-9846]{P.~Veres}
\affiliation{Department of Space Science, University of Alabama in Huntsville, Huntsville, AL 35899, USA}
\affiliation{Center for Space Plasma and Aeronomic Research, University of Alabama in Huntsville, Huntsville, AL 35899, USA}

\begin{abstract}
We explore fitting gamma-ray burst spectra with three physically-motivated models, and thus revisit the viability of synchrotron radiation as the primary source of GRB prompt emission. We pick a sample of 100 bright GRBs observed by the Fermi Gamma-ray Burst Monitor (GBM), based on their energy flux values. In addition to the standard empirical spectral models used in previous GBM spectroscopy catalogs, we also consider three physically-motivated models; (a) a Thermal Synchrotron model, (b) a Band model with a High-energy Cutoff, and (c) a Smoothly Broken Power Law (SBPL) model with a Multiplicative Broken Power Law (MBPL). We then adopt the Bayesian information criterion (BIC) to compare the fits obtained and choose the best model. We find that 42\% of the GRBs from the fluence spectra and 23\% of GRBs from the peak-flux spectra have one of the three physically-motivated models as their preferred one. From the peak-flux spectral fits, we find that the low-energy index distributions from the empirical model fits for long GRBs peak around the synchrotron value of -2/3, while the two low-energy indices from the SBPL+MBPL fits of long GRBs peak close to the -2/3 and -3/2 values expected for a synchrotron spectrum below and above the cooling frequency. 
\end{abstract}

\keywords{gamma rays: bursts --- methods: data analysis}

\section{Introduction} \label{sec:introduction}
The detailed nature of the radiative process responsible for gamma-ray burst (GRB) prompt emission has not yet been identified. The radiative process expected to dominate the emission is synchrotron radiation, due to the non-thermal appearance of the observed spectra and the likely presence of accelerated electrons and intense magnetic fields \citep{Katz_1994, Rees&Meszaros_1994, Tavani_1996}, but the inconsistency between the observed spectral shape at low energies and predictions from the synchrotron theory represents a challenge for this interpretation. In case of fast-cooling synchrotron radiation, the part of the spectrum immediately below the $\nu F_{\nu}$ peak energy should display a power-law behavior with a slope $\sim-3/2$, which breaks to a harder spectral shape (slope $\sim-2/3$) at lower energies \citep{Sari_1998}. The prompt emission spectra of GRBs are usually fit with empirical functions, such as the Band function \citep{Band_1993}, which consists of two smoothly connected power laws \textit{N(E)} $\propto$ $E^\alpha$ and \textit{N(E)} $\propto$ $E^\beta$, describing the photon spectrum at low and high energies, respectively. The typical slope of the low-energy power law is $\alpha\sim-1$ \citep{Gruber_2014, Goldstein_2012}. This is higher than the value expected in the case of fast-cooling synchrotron radiation.

Recent works have shown that a number of GRBs have an additional spectral break between $\sim$1 and a few hundred keV \citep{Oganesyan_2017,Ravasio_2019,Toffano_2021} and the slopes of the power-law below and above this break are consistent with the values expected from synchrotron emission. It has been suggested that the value of $\alpha$ is an average value between the two power-law segments below and above the break energy. The low-energy power-law index is of particular interest for determining the emission mechanism that converts the kinetic or/and magnetic energy of the bulk relativistic outflow into radiation. Under the assumption that the emission is dominated by synchrotron radiation, the low-energy photon index should be no harder than $-3/2$ in the case of non-adiabatic cooling, and no harder than $-2/3$ in the case of adiabatic cooling~\citep{RybickiLightman_1979, Katz_1994}.

In this work we explore three physically-motivated models to fit GRB Spectra, and thus revisit the viability of synchrotron radiation as the source of GRB prompt emission. We pick a sample of 100 bright GRBs observed by GBM, based on their energy flux values. In addition to the empirical models used in the latest Spectral Catalog \citep{Poolakkil_2021}, we also consider three new models; (a) a Thermal Synchrotron model, (b) a Band model with a high-energy cutoff, with the power-law indices fixed at the synchrotron values (-0.67 and -1.5 respectively), and (c) a Smoothly Broken Power-law (SBPL) model with a multiplicative broken power-law (MBPL). We then adopt the Bayesian information criterion \citep{Neath_Cavanaugh_2012} to compare the fits obtained and choose the best one.

\section{Analysis Method \label{sec:analysis method}}

\subsection{Instrument and Sample Selection}
{Fermi} GBM consists of 14 detector modules: 12 Sodium Iodide (NaI) detectors, covering the energies 8 - 1000 keV, and two Bismuth Germanate (BGO) detectors, covering 200 keV to 40 MeV \citep{Meegan_2009}. Data selection is identical to that as described in previous spectroscopy catalogs \citep{Gruber_2014,Poolakkil_2021}. In brief, up to three NaI detectors with observing angles to the source less than $60^{\circ}$ are selected, along with the BGO detector that has the smallest observing angle of the burst. For each of these, standard energy ranges that avoid unmodeled effects, such as an electronic roll-off at low energies and high-energy overflow bins are selected. Each data set is binned according to whether the burst is long (1.024 s binning) or short (0.064 s binning), with a dividing line at $T_{90}$ = 2 s \citep{Kouveliotou_1993}, where $T_{90}$ is the time between the 5\% and 95\% values of the total fluence. Next, a background model (polynomial in time) is chosen to fit regions of the light curve that bracket the emission interval.

The analyses presented herein are comprised of two spectra for each burst: a `fluence' (\textit{F}) spectrum that represents the entire duration of emission and a `peak flux' (\textit{P}) spectrum that depicts the brightest portion of each burst, on a fixed timescale of 1.024 s for long GRBs and 64 ms for short GRBs. The selection of fluence time bins for each of these two classes is made by including every (energy-integrated) time bin that has flux that is at least 3.5 sigma in excess of the background model for that bin. The data is then joint fit with RMfit (version 4.5.3\footnote{\url{https://fermi.gsfc.nasa.gov/ssc/data/analysis/user/}}, available at the Fermi Science Support Center), using a set of standard model functions (Section~\ref{sec:Models}). For a fit statistic, we have chosen the pgstat statistic \citep{Arnaud_2011}, which properly accounts for the Gaussian uncertainties in the background, arising from the temporal fit.

\begin{figure}
\gridline{\fig{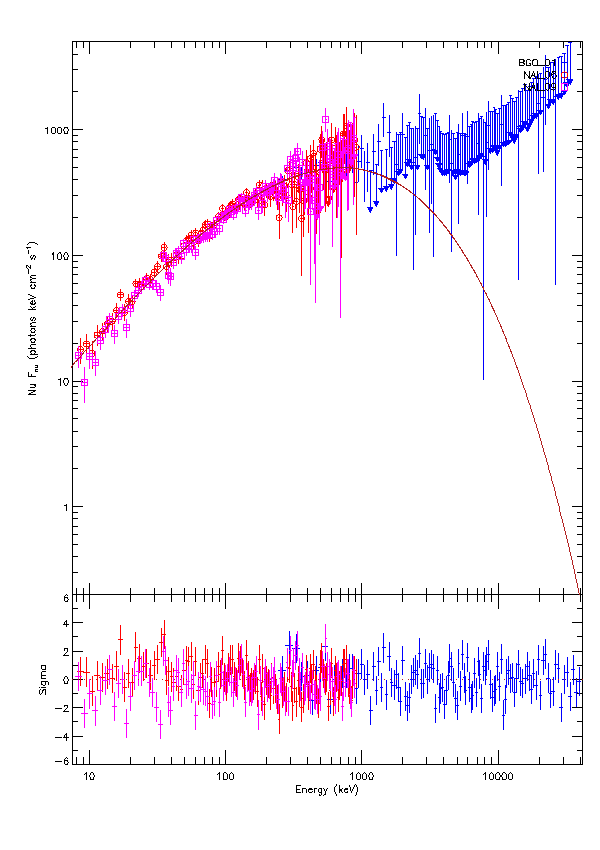}{0.33\textwidth}{(a)}\hspace{0mm}%
          \fig{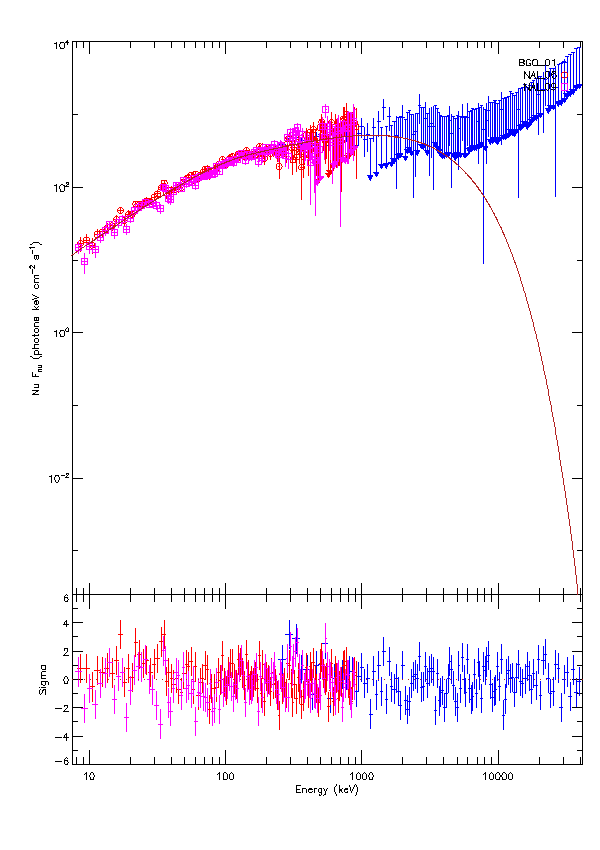}{0.33\textwidth}{(b)}\hspace{0mm}%
          \fig{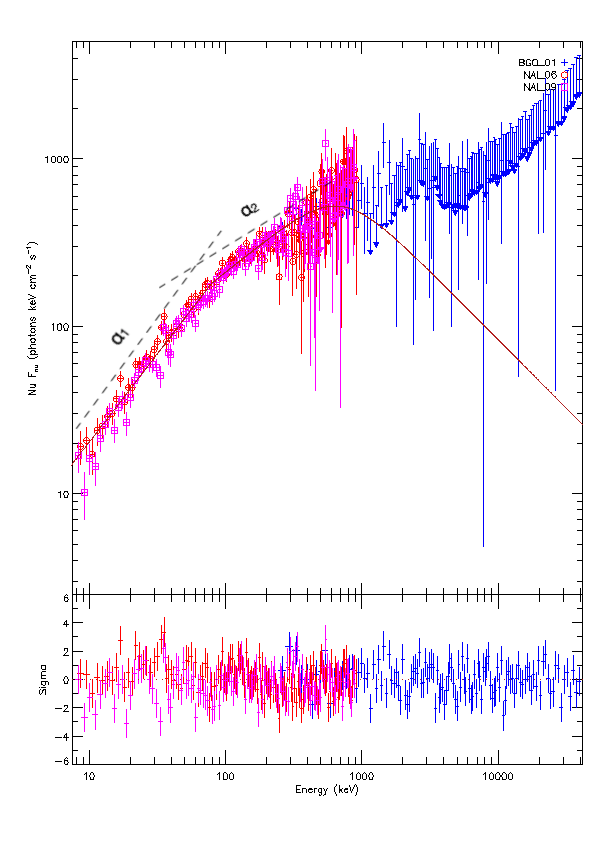}{0.33\textwidth}{(c)}}\hspace{0mm}% 
% \vspace*{-10mm}          
\caption{$\nu F_{\nu}$ (photons keV cm$^{-2}$ s$^{-1}$) spectrum for TS (\textit{left}), Band+H.E. Cutoff (\textit{middle}) and SBPL+MBPL (\textit{right}) fits in RMfit for GRB 110921912. The pink/blue data points represent data from the NaI/BGO detectors respectively and the red dashed line represents the spectral shape predicted by the model. }
\label{fig:Model_nuFnu}  
\end{figure}

\subsection{Models} \label{sec:Models}
We chose seven spectral models to fit the spectra of GRBs in our sample. In addition to the empirical spectral functions used in the latest spectroscopy catalog, we consider three additional models that are physically motivated \citep{Bauke_2007, Arnold_2015}. We also consider the COMP and BAND models with a Black Body component added to them. All models are formulated in units of photon flux with energy (\textit{E}), temperature (\textit{kT}) in keV and multiplied by a normalization constant \textit{A} (photon\ s$^{-1}$ cm$^{-2}$ keV$^{-1}$). Note that for the SBPL+MBPL model, the two low-energy indices before and after the MBPL break energy will be referred to as $\alpha_{1}$ and $\alpha_{2}$ respectively, as highlighted in Figure~\ref{fig:Model_nuFnu}(c). The pivot energy ($E_{\rm piv}$) normalizes the model to the energy range under consideration and helps reduce cross-correlation of other parameters. In all cases, $E_{\rm piv}$ is held fixed at 100 keV.

\begin{itemize}
\item \textit{COMP:} An exponentially-attenuated power law (`comptonized'), with normalization ($A$), low-energy power-law index ($\alpha$) and characteristic energy ($E_{\rm peak}$)
\begin{equation}
  f_{COMP}(E) = A \left( \frac{E}{E_{piv}} \right)^{\alpha} \exp \left[ -\frac{(\alpha + 2) E}{E_{peak}}  \right] \label{eq:comp}
\end{equation}

\item \textit{BAND:} The Band GRB function, with normalization ($A$), low-energy power-law index ($\alpha$), high-energy power-law index ($\beta$) and characteristic energy ($E_{\rm peak}$)
\begin{equation}
   f_{BAND}(E) = A \left\{
               \begin{array}{ll}
                 \left( \frac{E}{100 \ \rm keV} \right)^{\alpha} \exp \left[ -\frac{(\alpha + 2) E}{E_{peak}}  \right], & E \geq \frac{(\alpha - \beta) E_{peak}}{\alpha +2}    \\
                 \left( \frac{E}{100 \ \rm keV} \right)^{\beta} \exp\left(\beta - \alpha \right)\left[ \frac{(\alpha - \beta) E_{peak}}{100 \ \rm keV (\alpha +2)}  \right]^{\alpha - \beta}, & E < \frac{(\alpha - \beta) E_{peak}}{\alpha +2}
               \end{array}
             \right.
\end{equation}

\item \textit{SBPL:} A smoothly broken power law, with normalization ($A$), low-energy power-law index ($\lambda_1$), high-energy power-law index ($\lambda_2$), a characteristic break energy ($E_{\rm b}$) and the break scale ($\Delta$), in decades of energy. As in \citet{Gruber_2014}, we keep the value of $\Delta$ fixed at 0.3.
\begin{equation}
  f_{{\rm SBPL}}(E)=A \biggl (\frac{E}{E_{{\rm piv}}} \biggr)^b \ 10^{(a - a_{{\rm piv}})},
\end{equation}
where:
\begin{eqnarray*}
    a=m\Delta\ln\left(\frac{e^q+e^{-q}}{2}\right),\quad & \quad a_{\rm piv}=m\Delta\ln\left(\frac{e^{q_{\rm piv}}+e^{-q_{\rm piv}}}{2}\right),\\
    q=\frac{\log (E/E_b)}{\Delta},\quad & \quad q_{\rm piv}=\frac{\log (E_{\rm piv}/E_b)}{\Delta},\\
    m=\frac{\lambda_2-\lambda_1}{2},\quad & {\rm and} \quad b=\frac{\lambda_2+\lambda_1}{2}
\end{eqnarray*}

\item \textit{MBPL:} A Multiplicative Broken Power Law, which can be used as a low-energy or high-energy cutoff by fixing
$\lambda_h$ or $\lambda_l$ at zero, respectively
\begin{equation}
    f_{MBPL}(E) =\left\{ \begin{array}{cc}
     (E/E_b)^{\lambda_l},     &  E \leq E_b \\
     (E/E_b)^{\lambda_h} ,    & E > E_b
    \end{array}
    \right.
\end{equation}

\item \textit{Black Body:} A Black Body spectrum with normalization ($A$) and temperature (\textit{kT})
\begin{equation}
    f_{BB}(E) = A \left( \frac{E^2}{\exp(E/kT)-1} \right)
\end{equation}

\item \textit{High-energy Cutoff:} A multiplicative component with a cutoff energy ($E_{\rm cut}$) and folding energy ($E_{\rm F}$)
\begin{equation}
    f_{HEC}(E) =\left\{ \begin{array}{cc}
    1,     &  E \leq E_{\rm cut} \\
     (E/E_{\rm cut})^{(E_{\rm cut}/E_{\rm F})} \exp[\frac{(E_{\rm cut}-E)}{E_{\rm F}}],    & E > E_{\rm cut}
    \end{array}
    \right.
\end{equation}

\item \textit{Thermal Synchrotron:} For this model, we assume an electron distribution that consists of electrons in a thermal 
pool, given by
\begin{equation}
  n_{e}(\gamma) = n_{0} \left[ \left( \frac{\gamma}{\gamma_{th}} \right)^{2} \exp \left( -\frac{\gamma}{\gamma_{th}} \right) \right]
\end{equation}

Here $n_{0}$ normalizes the distribution to total number or energy, $\gamma$ is the electron Lorentz factor in the fluid frame
and $\gamma_{th}$ is the thermal electron Lorentz factor. We convolve this simplified distribution with the standard isotropic synchrotron kernel \citep{RybickiLightman_1979}
\begin{equation}
  F_{v}(\varepsilon) \propto \int_{1}^{\infty} n_{e}(\gamma) \mathcal{F} \left( \frac{\varepsilon}{\varepsilon_{c}} \right) d\gamma,
\end{equation}
where
\begin{equation}
  \mathcal{F}(w) = \int_{w}^{\infty} K_{5/3}(x)dx
\end{equation}
expresses the single-particle synchrotron emissivity (i.e., energy per unit time per unit volume) in dimensionless functional form.

\end{itemize}

\begin{figure}
\gridline{\fig{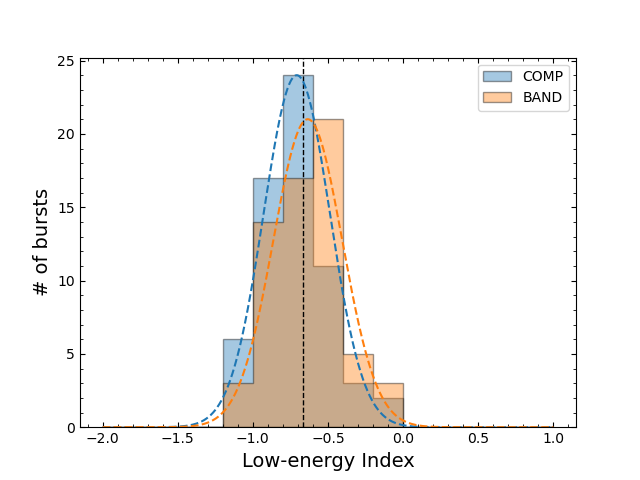}{0.5\textwidth}{(a)}
          \fig{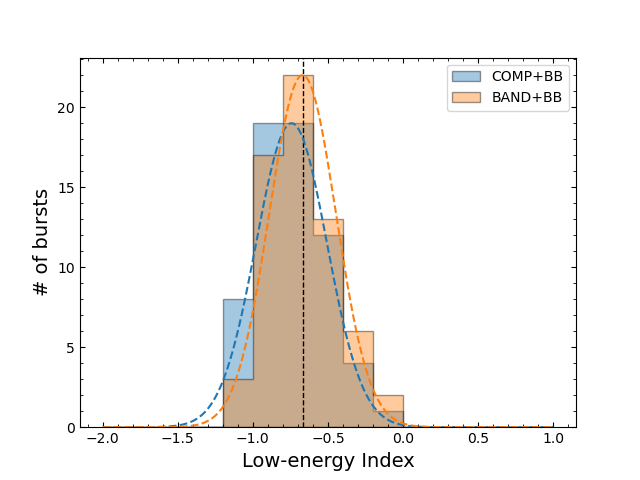}{0.5\textwidth}{(b)}}
\caption{Distribution of the low-energy indices from COMP, BAND, COMP+BB and BAND+BB fits for long GRBs (Peak-flux Spectra). Gaussian 
functions showing the central value and standard deviation of the distributions are overlapped to the histograms (colour-coded dashed 
curves). The black dashed-line represents the synchrotron ``line-of-death" value of -2/3.}
\label{fig:Empirical_LE}
\end{figure}

\begin{figure}
\gridline{\fig{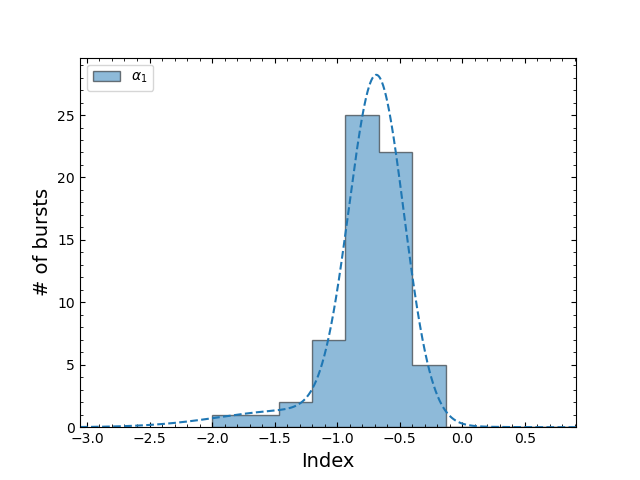}{0.5\textwidth}{(a)}
          \fig{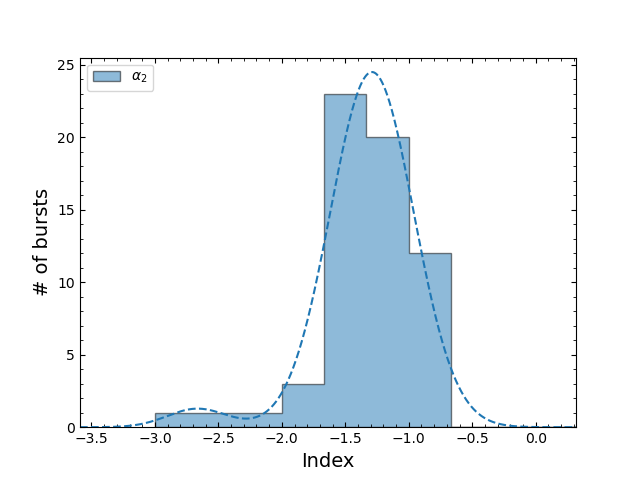}{0.5\textwidth}{(b)}}
\caption{Distribution of the two low-energy indices from the SBPL+MBPL fits for long GRBs (Peak-flux Spectra). Gaussian functions showing 
the central value and standard deviation of the distributions are overlapped to the histograms (dashed curves).}
\label{fig:SBPL_MBPL_LE}
\end{figure}

\subsection{PGSTAT and BIC}
In previous works we have used a variant of the Cash-statistic likelihood \citep{Cash_1979}, called C-Stat in RMfit and pstat in Xspec \citep{Arnaud_2011}, which assumes the background model uncertainty to be negligible. For this work, we have chosen to use pgstat, which correctly accounts for the Gaussian uncertainties in the temporally-interpolated background model, while retaining the Poisson statistics of the source counts.

We then adopt the Bayesian information criterion (BIC) to compare the fits obtained and choose the best one. The BIC is a well-known general approach to model selection that favors more parsimonious models over more complex models, i.e. it adds a penalty based on the number of parameters being estimated in the model. One form for calculating the BIC is given by:
\begin{equation}
    BIC = -2\:ln(\hat{L}) + k\:ln(n)
\end{equation}
where $\hat{L}$ is the maximized value of the likelihood function of the model $M_{k}$, \textit{k} is the number of free parameters in the model and \textit{n} is the number of data points. The model with the lowest BIC is considered the best. We can also calculate the $\Delta$ BIC; the difference between a particular model and the ‘best’ model with the lowest BIC, and use it as an argument against the other model. Applying these criteria, the number of bursts that classify as `best' for each model can be seen in Table~\ref{tab:BIC_Best}. We find that 73\% of Short GRBs in the fluence spectra and 84\% of Short GRBs in the peak-flux spectra have COMP as the preferred model. The low-energy index distributions from the empirical model fits for long GRBs are shown in Figure~\ref{fig:Empirical_LE}. Table~\ref{tab:Best_Models_BIC} shows that out of the 100 GRBs used in this sample, 42 in the fluence spectra and 23 in the peak-flux spectra have one of the three physically motivated models as their preferred one. A detailed list of the BIC values for all bursts, for both the fluence and peak-flux spectra can be found in Table~\ref{tab:BIC_Comparison_flnc} and Table~\ref{tab:BIC_Comparison_pflx} respectively.

\def\arraystretch{1.3}
\begin{table}[h]
    % \movetableright=-0.75in
    \caption{Best GRB Models based on BIC}
    \centering
    \begin{tabular}{l|c|c}
	\hline \hline
	~ & Fluence Spectra & Peak-flux Spectra\\ \hline
	Empirical Models   & 58 & 77  \\
	Physically-Motivated Models & 42 & 23  \\ \hline 
    \end{tabular}
    \label{tab:Best_Models_BIC}
\end{table}

\begingroup
\begin{table}
    \movetableright=-0.85in
    \caption{BIC Comparison (Number of GRBs for which the given model has the lowest value of BIC)}
    \centering
    \begin{tabular}{l|c|c|c|c|c|c}
	\hline \hline
	Model & Total & Total & Long GRBs & Short GRBs & Long GRBs & Short GRBs \\
	~ & (fluence) & (peak-flux) & (fluence) & (fluence) & (peak-flux) & (peak-flux) \\
	 \hline
	TS   & 9 & 2 & 4 & 5 & 2 & 0\\
	COMP & 31 & 41 & 4 & 27 & 10 & 31\\ 
	COMP+BB & 2 & 9 & 2 & 0 & 7 & 2\\
	BAND & 11 & 15 & 10 & 1 & 13 & 2\\
	BAND+BB & 14 & 12 & 14 & 0 & 11 & 1 \\
	BAND+H.E. Cutoff & 12 & 14 & 9 & 3 & 13 & 1 \\
	SBPL+MBPL & 21 & 7 & 20 & 1 & 7 & 0 \\
	\hline
    \hline 
    \end{tabular}
    \label{tab:BIC_Best}
\end{table}
\endgroup

\section{Parameter Distributions} \label{sec:parameter distrubutions}
The distribution of spectral parameters help us place each burst in relation to the ensemble of all bursts. The time-integrated spectral distributions depict the overall emission properties and any spectral evolution is averaged out. For peak spectra however, the time intervals are reasonably short for any significant change of the spectral parameters.

A single electron emitting synchrotron radiation will have a spectra with slope -2/3 (photon index) and an exponential cutoff at high energies. The spectrum of a population of radiating electrons in the optically thin regime is just a superposition of individual spectra. For this reason the -2/3 value is also a hard upper limit on the overall spectral slope, also referred to as the synchrotron “line of death” \citep{Preece_1998a}. For a population of shock accelerated electrons,  the part of the spectrum immediately below the $\nu F_{\nu}$ peak energy will display a power-law behavior with a slope $\sim-3/2$, which breaks to a harder spectral shape (slope $\sim-2/3$, following the single electron spectrum) at lower energies. This is in the so called fast-cooling regime, where random Lorentz factor of an electron that cools on the dynamic timescale, is lower than the typical injected electron's energy \citep{Sari_1998}. 

Figure~\ref{fig:Empirical_LE} shows the distribution of low-energy indices of long GRBs (Peak-flux spectra) from the four empirical models used in this work. They have a combined median value of $\sim$ $-0.69_{-0.22}^{+0.20}$, aligning with the $-2/3$ expectation. The distribution of the two low-energy indices from the SBPL+MBPL fits for peak-flux spectra of long GRBs are shown in Figure~\ref{fig:SBPL_MBPL_LE}; the two distributions have median values of $-0.72_{-0.21}^{+0.20}\, (\alpha_1)$ and $-1.32_{-0.24}^{+0.33}\, (\alpha_2)$ respectively (Table~\ref{tab:Short and Long}). These values are in line with the -2/3 and -3/2 expected indices for synchrotron spectrum from fast-cooling electrons. This result lends strong support for the case of synchrotron radiation as the primary emission mechanism for GRB prompt emission. 

\begin{figure}
\gridline{\fig{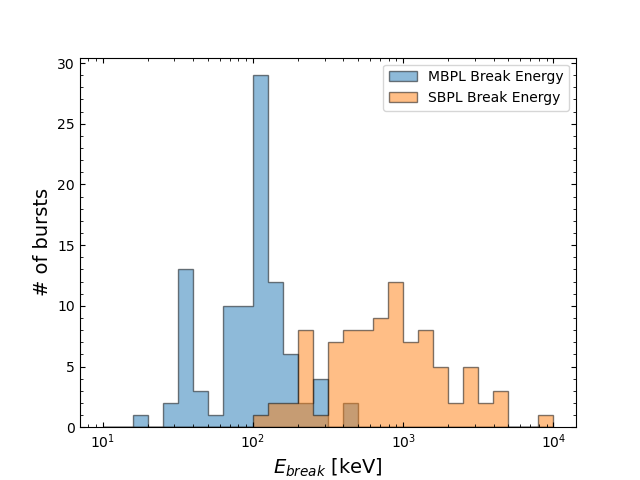}{0.5\textwidth}{(a)}
          \fig{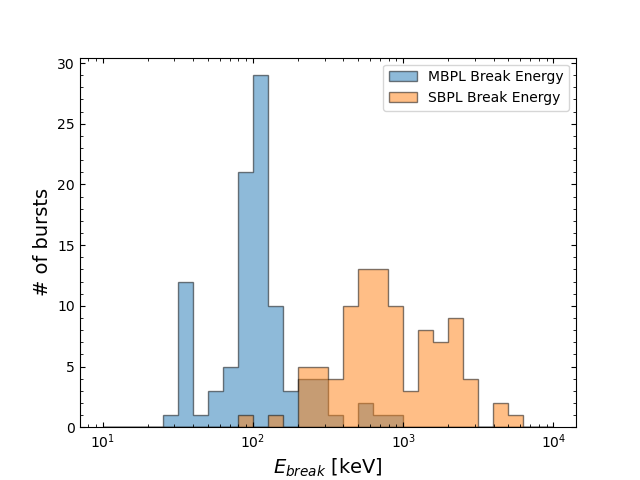}{0.5\textwidth}{(b)}}
\caption{Distribution of the two break energies from the fluence (a) and peak-flux (b) spectral fits for SBPL+MBPL.}
\label{fig:SBPL_MBPL_Ebreak}
\end{figure}

\begingroup
\def\arraystretch{1.3}
\begin{table}[h]
    \movetableright=-0.75in
    \caption{The median parameter values and the 68\% CL for all long and short GRBs}
    \centering
    \begin{tabular}{l|ccc|ccc}
	\hline \hline
	~    & ~                       & Long GRBs               & ~                  & ~                       & Short GRBs              & ~                  \\ \hline
	Model & L.E. Index & H.E. Index & $E_{peak}$ (keV) & L.E. Index & H.E. Index & $E_{peak}$ (keV) \\
	 \hline
	 \multicolumn{7}{c}{Fluence Spectra}\\
	 \hline
    COMP & $-0.97_{-0.20}^{+0.20}$ & ... & $372_{-162}^{+412}$ &$-0.58_{-0.30}^{+0.28}$ & ... & $1024_{-512}^{+331}$ \\
    COMP+BB & $-0.99_{-0.21}^{+0.21}$ & ... & $434_{-183}^{+477}$ &$-0.65_{-0.23}^{+0.24}$ & ... & $982_{-477}^{+516}$ \\
    BAND & $-0.91_{-0.18}^{+0.19}$ & $-2.41_{-0.39}^{+0.23}$ & $308_{-143}^{+473}$ &$-0.54_{-0.27}^{+0.26}$ & $-3.28_{-7.93}^{+1.09}$ & $967_{-516}^{+371}$ \\
    BAND+BB & $-0.93_{-0.19}^{+0.27}$ & $-2.55_{-0.80}^{+0.32}$ & $378_{-226}^{+486}$ & $-0.55_{-0.27}^{+0.19}$ & $-3.18_{-5.21}^{+1.05}$ & $903_{-482}^{+471}$ \\
    SBPL+MBPL & $-0.97_{-0.30}^{+0.23}$, $-1.41_{-0.48}^{+0.27}$ & $-2.61_{-0.46}^{+0.44}$ & $497_{-290}^{+470}$ & $-0.59_{-0.32}^{+0.35}$, $-0.94_{-0.45}^{+0.24}$ & $-3.15_{-3.59}^{+0.91}$ & $1074_{-496}^{+584}$ \\
	\hline
	\multicolumn{7}{c}{Peak Flux Spectra}\\
	\hline
    COMP & $-0.71_{-0.19}^{+0.20}$ & ... & $423_{-164}^{+803}$ & $-0.27_{-0.36}^{+0.29}$ & ... & $830_{-318}^{+509}$ \\
    COMP+BB & $-0.74_{-0.22}^{+0.24}$ & ... & $468_{-132}^{+739}$ & $-0.46_{-0.27}^{+0.37}$ & ... & $857_{-300}^{+606}$ \\
    BAND & $-0.60_{-0.22}^{+0.14}$ & $-2.66_{-1.06}^{+0.40}$ & $377_{-161}^{+770}$ & $-0.22_{-0.39}^{+0.37}$ & $-4.80_{-6.06}^{+2.41}$ & $834_{-385}^{+500}$ \\
    BAND+BB & $-0.68_{-0.20}^{+0.22}$ & $-2.91_{-1.11}^{+0.53}$ & $468_{-199}^{+471}$ & $-0.39_{-0.37}^{+0.61}$ & $-4.58_{-4.70}^{+2.13}$ & $895_{-406}^{+519}$ \\
    SBPL+MBPL & $-0.72_{-0.21}^{+0.20}$, $-1.32_{-0.24}^{+0.33}$ & $-3.11_{-1.34}^{+0.67}$ & $493_{-162}^{+694}$ & $-0.30_{-0.36}^{+0.62}$, $-0.89_{-0.28}^{+0.45}$ & $-4.30_{-5.90}^{+1.59}$ & $905_{-342}^{+533}$ \\ 
    \hline 
    \end{tabular}
    \label{tab:Short and Long}
\end{table}
\endgroup

It is evident from Figure~\ref{fig:Empirical_LE}, \ref{fig:SBPL_MBPL_LE} that a non-negligible fraction of GRBs have a low-energy index value steeper than -2/3, which presents a challenge for the synchrotron theory. One possibility in such cases would be the appearance of a quasi-thermal component of photospheric origin \citep{Meszaros&Rees_2000,Daigne_2002}. The emission from both the photosphere and internal shocks have a similar duration, the latter having only a very short lag behind the first. The intensity of the photospheric emission depends strongly on the unknown mechanism responsible for the acceleration of the relativistic outflow \citep{Daigne_2011}. Another possibility to consider is the effect of electron pitch angle scattering \citep{LLoyd&Petrosian_2000}; where a small pitch angle leads to a harder low-energy index value ($-2/3 \: \lesssim \: \alpha \: \lesssim 0$) and a large pitch angle can accommodate bursts below the “line of death” ($\alpha \lesssim -2/3$). Spectral indices harder than -3/2 can be obtained by more elaborate models. For example, \cite{Derishev_2007, Zhao_2014} showed that fast cooling electrons in decaying magnetic fields can lead to such hard spectra.

Figure \ref{fig:SBPL_MBPL_Ebreak} shows the distribution of the two break energies from the SBPL+MBPL model. Interestingly, we find a subgroup of GRBs that have a low-energy break between 30-40 keV in both the fluence and peak-flux spectral fits. A similar low-energy break was found in GRB 221009A \citep{Lesage_2023} and will be investigated further in a future work. In the scenario where a low-energy spectral break is a common feature in GRB spectra, it has been suggested that the average low-energy spectral index value of $\alpha \sim$ -1 \citep{Gruber_2014,Poolakkil_2021} is an average value between the two power-law segments below and above the break energy \citep{Toffano_2021}. We find that for our sample of bursts, only long GRBs from the peak-flux spectra satisfy this condition (Table~\ref{tab:Short and Long}).

\section{Summary} \label{sec:conclusions}
The goal of this work is to explore the the use of physically-motivated models to fit GRB spectra and directly compare them to the empirical models used in previous GBM spectroscopy catalogs \citep{Goldstein_2012,Gruber_2014}. The spectral properties presented here are from time-integrated and peak flux analysis, produced using seven photon models; four empirical and three physically-motivated. We use a sample of 100 GRBs, that were selected based on their energy flux values \citep{Poolakkil_2021}. We have used pgstat as our fitting statistic; which accounts for the Gaussian uncertainties in the background model, while retaining the Poisson statistics of the source counts, and then use the Bayesian information criterion (BIC) to compare fits and choose the best model.

We found that 42\% of the GRBs from the fluence spectra and 23\% of GRBs from the peak-flux spectra had one of the three physically-motivated models as their preferred one. The low-energy index distributions from the empirical model fits for long GRBs (peak-flux spectra), highlighted in Figure~\ref{fig:Empirical_LE}, peak around the synchrotron value of -2/3, while the two low-energy indices from the SBPL+MBPL fits of long GRBs (peak-flux spectra, Figure~\ref{fig:SBPL_MBPL_LE}) peak close to the -2/3 and -3/2 values expected for synchrotron spectrum below and above the cooling frequency. These results present a strong case for synchrotron radiation as a leading mechanism for the origin of prompt emission from GRBs and further encourage the transition from empirical models towards physically-motivated models to fit GRB spectra.

The UAH co-authors gratefully acknowledge NASA funding from co-operative agreement 80MSFC22M0004.

\startlongtable
\begin{deluxetable}{lcccccccc}

\tabletypesize{\footnotesize}
\tablecaption{BIC Comparison (Fluence Spectra). The best model is indicated with an $^{*}$ \label{tab:BIC_Comparison_flnc}}
 
\tablehead{\colhead{ID} & \colhead{TS } & \colhead{COMP} & \colhead{COMP+BB} & \colhead{BAND} & \colhead{BAND+BB} & \colhead{BAND+H.E.C} & \colhead{SBPL+MBPL}} 
\startdata
&\multicolumn{7}{c}{Short GRBs} \\  \cline{2-8}
GRB 170305256 & 520.29 & 502.81$^{*}$ & 514.29 & 512.22 & 523.76 & 503.21 & 524.21 \\
GRB 100223110 & 402.04 & 343.67$^{*}$ & 353.61 & 349.59 & 359.48 & 366.13 & 357.25 \\
GRB 120817168 & 551.69 & 549.77$^{*}$ & 562.02 & 555.96 & 568.16 & 553.63 & 566.14 \\
GRB 110529034 & 458.97$^{*}$ & 468.32 & 480.98 & 470.75 & 483.07 & 469.75 & 483.5 \\
GRB 180402406 & 582.6 & 531.31$^{*}$ & 531.52 & 537.72 & 546.1 & 543.79 & 537.54 \\
GRB 120624309 & 558.15 & 523.56 & 523.51 & 529.21 & 527.84 & 598.61 & 523.36$^{*}$ \\
GRB 130701761 & 533.26 & 432.26$^{*}$ & 434.34 & 438.36 & 440.44 & 438.35 & 445.76 \\
GRB 170219002 & 231.8$^{*}$ & 238.44 & 244.87 & 244.28 & 255.49 & 236.75 & 250.86 \\
GRB 140209313 & 535.41 & 442.16 & 423.7 & 412.69 & 413.99 & 408.59$^{*}$ & 446.9 \\
GRB 090227772 & 836.32 & 545.88$^{*}$ & 552.22 & 549.3 & 555.26 & 582.4 & 548.49 \\
GRB 170816599 & 568.19 & 501.91$^{*}$ & 510.62 & 508.09 & 516.81 & 503.97 & 517.1 \\
GRB 170708046 & 490.4$^{*}$ & 503.12 & 505.58 & 501.34 & 513.16 & 492.82 & 512.75 \\
GRB 130804023 & 518.35 & 515.33$^{*}$ & 527.55 & 521.25 & 532.61 & 515.38 & 533.82 \\
GRB 130628860 & 493.79 & 487.86 & 499.97 & 494.07 & 506.13 & 487.79$^{*}$ & 506.11 \\
GRB 100811108 & 627.64 & 504.62$^{*}$ & 513.52 & 511.31 & 520.4 & 551.91 & 520.72 \\
GRB 091012783 & 413.61 & 384.0$^{*}$ & 395.3 & 389.87 & 401.61 & 387.56 & 400.66 \\
GRB 180204109 & 510.18$^{*}$ & 514.89 & 526.55 & 521.87 & 533.69 & 518.83 & 534.71 \\
GRB 090510016 & 487.08 & 484.15$^{*}$ & 493.15 & 486.64 & 492.85 & 504.56 & 494.15 \\
GRB 110705151 & 629.85 & 483.33$^{*}$ & 483.68 & 490.84 & 491.16 & 533.44 & 495.08 \\
GRB 170127067 & 607.27 & 238.33$^{*}$ & 248.93 & 242.76 & 253.78 & 379.68 & 249.58 \\
GRB 121127914 & 410.24 & 397.18$^{*}$ & 408.26 & 403.11 & 492.63 & 397.34 & 412.5 \\
GRB 150922234 & 522.61 & 514.68$^{*}$ & 525.04 & 520.77 & 532.42 & 515.08 & 532.05 \\
GRB 140901821 & 790.89 & 551.71$^{*}$ & 556.49 & 557.96 & 562.67 & 604.92 & 568.96 \\
GRB 111222619 & 523.8 & 397.59$^{*}$ & 409.44 & 403.48 & 415.12 & 413.62 & 415.21 \\
GRB 161218222 & 553.44 & 503.42$^{*}$ & 515.4 & 509.44 & 521.57 & 515.72 & 518.04 \\
GRB 150819440 & 480.21 & 444.13$^{*}$ & 447.56 & 450.04 & 453.44 & 512.65 & 456.33 \\
GRB 150811849 & 502.21 & 377.0$^{*}$ & 386.21 & 382.58 & 392.16 & 409.67 & 395.1 \\
GRB 130515056 & 390.97 & 375.3$^{*}$ & 385.82 & 381.47 & 392.74 & 379.17 & 393.3 \\
GRB 100206563 & 563.52 & 554.07$^{*}$ & 566.35 & 558.68 & 568.43 & 557.63 & 568.93 \\
GRB 131126163 & 421.71 & 370.5 & 379.11 & 369.03$^{*}$ & 380.78 & 392.54 & 381.77 \\
GRB 180703949 & 762.43 & 572.08 & 574.34 & 573.96 & 580.27 & 569.56$^{*}$ & 650.02 \\
GRB 081216531 & 557.73 & 555.17$^{*}$ & 567.51 & 560.93 & 566.42 & 555.48 & 560.26 \\
GRB 081209981 & 354.35 & 349.15$^{*}$ & 359.72 & 356.22 & 366.87 & 350.51 & 367.4 \\
GRB 141011282 & 489.65 & 482.31$^{*}$ & 494.52 & 488.46 & 500.68 & 482.39 & 500.72 \\
GRB 150118927 & 531.28$^{*}$ & 543.08 & 549.77 & 546.12 & 553.75 & 536.66 & 556.97 \\
GRB 120811014 & 456.22 & 381.04$^{*}$ & 391.21 & 386.93 & 397.09 & 402.99 & 399.36 \\
GRB 090617208 & 495.44 & 485.2$^{*}$ & 497.15 & 492.19 & 504.32 & 489.95 & 504.1 \\
\hline
\hline
&\multicolumn{7}{c}{Long GRBs} \\  \cline{2-8}
GRB 090424592 & 943.46 & 904.7 & 908.36 & 896.75$^{*}$ & 902.77 & 974.74 & 907.22 \\
GRB 160802259 & 526.33 & 396.64 & 346.75 & 375.57 & 352.17 & 325.22$^{*}$ & 359.68 \\
GRB 160422499 & 965.24$^{*}$ & 1088.0 & 997.55 & 1018.5 & 1002.31 & 1109.41 & 988.35 \\
GRB 131028076 & 980.67 & 595.22 & 591.52 & 590.07 & 597.68 & 612.87 & 583.79$^{*}$ \\
GRB 131014215 & 6203.19 & 1389.08 & 1059.02 & 1130.08 & 1028.82$^{*}$ & 1494.34 & 1042.63 \\
GRB 081009140 & 683.84 & 675.21$^{*}$ & 687.56 & 677.42 & 689.84 & 2876.54 & 945.1 \\
GRB 140206275 & 3086.79 & 1785.7 & 1697.75 & 1732.73 & 1678.66$^{*}$ & 1833.17 & 1696.65 \\
GRB 140329295 & 845.54 & 1009.1 & 793.84 & 810.49 & 724.86 & 701.44 & 692.03$^{*}$ \\
GRB 180218635 & 465.24 & 375.01 & 345.46 & 318.97 & 311.01$^{*}$ & 317.28 & 313.66 \\
GRB 110921912 & 523.83$^{*}$ & 567.53 & 554.52 & 562.72 & 556.55 & 550.56 & 560.53 \\
GRB 130606497 & 3282.79 & 2573.84 & 2327.23 & 2166.65 & 2048.68$^{*}$ & 2345.2 & 2199.81 \\
GRB 150201574 & 1215.08 & 1179.7 & 1103.17 & 1098.89 & 1087.78$^{*}$ & 1127.39 & 1359.07 \\
GRB 150118409 & 1000.63 & 826.55$^{*}$ & 829.55 & 831.46 & 834.55 & 1428.47 & 840.13 \\
GRB 090618353 & 2268.14 & 2417.27 & 2299.55 & 2183.23$^{*}$ & 2194.57 & 2300.75 & 2240.15 \\
GRB 171010792 & 6546.13 & 8247.58 & 6841.33 & 6932.89 & 5208.78 & 5626.73 & 5030.72$^{*}$ \\
GRB 180113418 & 2136.99 & 2424.76 & 1607.83 & 1743.1 & 1457.65 & 1535.91 & 1317.56$^{*}$ \\
GRB 150902733 & 1619.4 & 945.88 & 772.38 & 808.85 & 743.54$^{*}$ & 840.72 & 754.96 \\
GRB 180113011 & 697.05 & 691.22$^{*}$ & 703.06 & 694.66 & 706.03 & 744.81 & 704.02 \\
GRB 100701490 & 512.2 & 523.73 & 527.83 & 501.63$^{*}$ & 509.23 & 535.97 & 511.47 \\
GRB 170409112 & 1432.47 & 1462.62 & 1304.64 & 1365.75 & 1326.63 & 1752.57 & 1256.37$^{*}$ \\
GRB 120426090 & 772.78 & 526.49 & 500.81 & 516.3 & 505.27 & 493.39$^{*}$ & 499.03 \\
GRB 140508128 & 585.67 & 548.09 & 558.34 & 529.48$^{*}$ & 530.49 & 600.19 & 554.45 \\
GRB 150330828 & 1703.4 & 1804.39 & 1702.53 & 1643.69$^{*}$ & 1657.5 & 1809.89 & 1657.99 \\
GRB 100719989 & 811.92 & 769.56 & 738.87 & 734.26 & 732.43 & 720.68$^{*}$ & 734.11 \\
GRB 131229277 & 635.12 & 511.93$^{*}$ & 519.3 & 518.0 & 522.26 & 526.88 & 520.86 \\
GRB 180305393 & 1361.37 & 773.62 & 722.06 & 734.21 & 721.81$^{*}$ & 793.52 & 722.52 \\
GRB 170115743 & 1702.69$^{*}$ & 1725.64 & 1708.17 & 1722.08 & 1713.56 & 1731.59 & 1724.21 \\
GRB 170527480 & 1475.9 & 947.69 & 906.9 & 953.82 & 913.11 & 1659.17 & 891.5$^{*}$ \\
GRB 110825102 & 840.33 & 806.61 & 795.5 & 804.54 & 800.68 & 958.97 & 795.27$^{*}$ \\
GRB 130502327 & 1797.47 & 1108.26 & 1120.67 & 1046.05$^{*}$ & 1055.15 & 1059.01 & 1058.96 \\
GRB 101014175 & 2518.6 & 2454.94 & 2325.33 & 2211.06 & 2222.85 & 2469.62 & 2208.61$^{*}$ \\
GRB 150510139 & 1108.36 & 734.44 & 705.58$^{*}$ & 740.59 & 711.72 & 1437.97 & 710.12 \\
GRB 161218356 & 3091.88 & 951.0 & 938.05 & 920.17$^{*}$ & 930.9 & 951.09 & 949.41 \\
GRB 120129580 & 576.46 & 501.91 & 435.06 & 484.29 & 431.59 & 437.62 & 419.8$^{*}$ \\
GRB 110625881 & 1378.57 & 1330.78 & 1262.54 & 1232.64 & 1171.76$^{*}$ & 1236.37 & 1255.76 \\
GRB 100826957 & 1976.3 & 2401.01 & 2098.61 & 1881.26 & 1675.54 & 1839.42 & 1669.4$^{*}$ \\
GRB 130305486 & 859.9 & 823.28 & 835.61 & 818.15 & 830.55 & 804.69$^{*}$ & 819.51 \\
GRB 170808936 & 1270.6 & 1247.67 & 1164.34 & 1171.63 & 1157.08 & 1509.21 & 1131.69$^{*}$ \\
GRB 160625945 & 4789.32 & 4819.52 & 4194.59 & 4414.85 & 4032.48 & 5500.55 & 4023.77$^{*}$ \\
GRB 090926181 & 4999.25 & 4640.18 & 4570.88 & 4581.35 & 4525.12 & 8281.4 & 4513.64$^{*}$ \\
GRB 140416060 & 384.73$^{*}$ & 444.95 & 393.13 & 408.23 & 408.19 & 431.21 & 482.19 \\
GRB 110301214 & 1193.92 & 1090.62 & 1009.64 & 1001.04 & 958.07$^{*}$ & 965.81 & 1291.02 \\
GRB 151231443 & 647.94 & 678.06 & 639.99 & 609.65 & 590.24 & 607.29 & 577.96$^{*}$ \\
GRB 130518580 & 1198.89 & 1208.92 & 1193.74 & 1177.23 & 1156.13$^{*}$ & 1230.43 & 1158.63 \\
GRB 091003191 & 635.59 & 626.53 & 636.14 & 623.81$^{*}$ & 635.59 & 676.12 & 637.77 \\
GRB 150403913 & 990.77 & 1078.27 & 996.93 & 1026.15 & 982.3 & 994.78 & 979.18$^{*}$ \\
GRB 090820027 & 2197.92 & 998.4 & 863.67 & 878.55 & 860.08 & 828.7$^{*}$ & 856.26 \\
GRB 090902462 & 4282.2 & 2398.21 & 1676.52 & 2404.35 & 1682.67 & 6148.28 & 1498.67$^{*}$ \\
GRB 171119992 & 323.57 & 334.03 & 345.01 & 328.48 & 335.33 & 321.69$^{*}$ & 334.31 \\
GRB 160910722 & 1014.37 & 1070.07 & 974.77 & 970.76 & 955.21 & 953.62 & 952.05$^{*}$ \\
GRB 120711115 & 1166.6 & 1023.68 & 1035.34 & 982.27$^{*}$ & 986.11 & 1506.99 & 987.48 \\
GRB 100829876 & 397.55 & 416.83 & 403.81 & 401.25 & 408.6 & 392.9$^{*}$ & 408.59 \\
GRB 150213001 & 1027.48 & 1187.29 & 1076.79 & 1066.79 & 972.23$^{*}$ & 1430.02 & 1313.7 \\
GRB 101123952 & 1097.33 & 1203.98 & 1175.41 & 1109.9 & 1082.61 & 1177.54 & 1076.64$^{*}$ \\
GRB 081215784 & 797.17 & 833.56 & 773.92 & 763.74 & 739.82 & 724.31$^{*}$ & 748.55 \\
GRB 131231198 & 2387.55 & 2229.76 & 2124.67 & 2088.08 & 2020.39$^{*}$ & 2131.04 & 2136.43 \\
GRB 160720767 & 1204.47 & 1195.51 & 1173.47$^{*}$ & 1183.06 & 1176.26 & 1300.46 & 1189.69 \\
GRB 150627183 & 1295.11 & 1522.15 & 1414.78 & 1327.61 & 1289.08$^{*}$ & 1482.12 & 1394.34 \\
GRB 160530667 & 2133.07 & 1109.48 & 986.9 & 952.01 & 940.4 & 916.17$^{*}$ & 1042.49 \\
GRB 140102887 & 671.8 & 605.1 & 614.73 & 601.09$^{*}$ & 610.02 & 625.96 & 662.68 \\
GRB 160509374 & 1347.4 & 1383.35 & 1396.4 & 1309.6 & 1321.38 & 1633.8 & 1300.26$^{*}$ \\
GRB 160821857 & 4327.84 & 2844.16 & 2650.69 & 2429.59 & 2335.56$^{*}$ & 4864.24 & 2362.25 \\
GRB 150210935 & 755.17 & 566.09 & 538.64 & 572.27 & 544.77 & 849.58 & 535.46$^{*}$ \\
\enddata

\end{deluxetable}

\startlongtable
\begin{deluxetable}{lcccccccc}

\tabletypesize{\footnotesize}
\tablecaption{BIC Comparison (Peak-flux Spectra). The best model is indicated with an $^{*}$ \label{tab:BIC_Comparison_pflx}}
 
\tablehead{\colhead{ID} & \colhead{TS } & \colhead{COMP} & \colhead{COMP+BB} & \colhead{BAND} & \colhead{BAND+BB} & \colhead{BAND+H.E.C} & \colhead{SBPL+MBPL}} 
\startdata
&\multicolumn{7}{c}{Short GRBs} \\  \cline{2-8}
GRB 170305256 & 501.08 & 470.17$^{*}$ & 476.24 & 480.2 & 485.45 & 480.53 & 492.58 \\
GRB 100223110 & 410.17 & 319.05$^{*}$ & 340.22 & 324.91 & 335.87 & 373.48 & 333.53 \\
GRB 120817168 & 546.63 & 535.22$^{*}$ & 547.73 & 541.82 & 548.18 & 535.3 & 548.84 \\
GRB 110529034 & 495.51 & 491.57$^{*}$ & 494.07 & 498.41 & 500.9 & 505.29 & 500.84 \\
GRB 180402406 & 561.08 & 530.84$^{*}$ & 543.17 & 538.32 & 549.31 & 546.18 & 546.13 \\
GRB 120624309 & 615.61 & 582.02 & 573.65$^{*}$ & 588.35 & 580.42 & 583.49 & 580.7 \\
GRB 130701761 & 469.61 & 454.73$^{*}$ & 461.21 & 460.83 & 467.3 & 461.13 & 467.1 \\
GRB 170219002 & 214.25 & 207.95$^{*}$ & 218.32 & 213.79 & 224.03 & 236.75 & 223.64 \\
GRB 140209313 & 461.14 & 391.15 & 380.19$^{*}$ & 382.94 & 384.87 & 395.59 & 384.92 \\
GRB 090227772 & 1194.71 & 510.8 & 516.37 & 507.09$^{*}$ & 515.1 & 878.5 & 514.75 \\
GRB 170816599 & 579.56 & 481.08$^{*}$ & 491.56 & 487.37 & 497.94 & 530.51 & 498.09 \\
GRB 170708046 & 484.0 & 490.02 & 482.41 & 475.69$^{*}$ & 487.86 & 479.13 & 486.31 \\
GRB 130804023 & 648.78 & 568.4$^{*}$ & 577.7 & 653.97 & 605.72 & 596.1 & 586.25 \\
GRB 130628860 & 541.73 & 508.59 & 519.32 & 514.71 & 506.08$^{*}$ & 520.96 & 506.11 \\
GRB 100811108 & 585.16 & 539.34$^{*}$ & 548.02 & 546.25 & 554.59 & 564.0 & 551.46 \\
GRB 091012783 & 350.11 & 329.4$^{*}$ & 340.96 & 335.51 & 346.92 & 335.91 & 345.85 \\
GRB 180204109 & 505.56 & 483.92$^{*}$ & 495.29 & 490.29 & 502.63 & 488.38 & 501.56 \\
GRB 090510016 & 460.4 & 442.46$^{*}$ & 452.11 & 449.07 & 458.43 & 442.84 & 460.25 \\
GRB 110705151 & 571.1 & 484.53$^{*}$ & 492.66 & 491.47 & 499.64 & 519.72 & 501.19 \\
GRB 170127067 & 520.46 & 248.25$^{*}$ & 253.89 & 253.75 & 266.19 & 369.71 & 260.91 \\
GRB 121127914 & 353.51 & 339.71$^{*}$ & 408.25 & 403.11 & 349.61 & 343.99 & 355.88 \\
GRB 150922234 & 493.62 & 461.22$^{*}$ & 472.98 & 467.42 & 479.16 & 472.89 & 478.11 \\
GRB 140901821 & 554.9 & 497.2$^{*}$ & 508.81 & 503.37 & 514.99 & 510.45 & 514.46 \\
GRB 111222619 & 424.32 & 414.79$^{*}$ & 426.39 & 421.51 & 431.56 & 415.14 & 431.17 \\
GRB 161218222 & 548.71 & 531.44$^{*}$ & 543.12 & 537.62 & 549.38 & 536.9 & 546.84 \\
GRB 150819440 & 733.35 & 402.54$^{*}$ & 406.27 & 408.43 & 405.89 & 538.07 & 403.79 \\
GRB 150811849 & 384.15 & 376.43$^{*}$ & 382.79 & 382.36 & 388.68 & 377.49 & 392.1 \\
GRB 130515056 & 377.85 & 363.33$^{*}$ & 374.88 & 369.14 & 380.42 & 370.07 & 379.39 \\
GRB 100206563 & 559.86 & 545.32$^{*}$ & 553.89 & 550.05 & 560.42 & 551.82 & 558.53 \\
GRB 131126163 & 376.45 & 333.25$^{*}$ & 343.46 & 336.31 & 347.49 & 361.01 & 347.56 \\
GRB 180703949 & 551.25 & 423.08$^{*}$ & 433.2 & 429.22 & 439.47 & 430.14 & 439.5 \\
GRB 081216531 & 563.22 & 518.73$^{*}$ & 530.56 & 525.44 & 537.05 & 529.24 & 533.2 \\
GRB 081209981 & 357.69 & 333.92$^{*}$ & 343.64 & 340.9 & 354.62 & 342.87 & 352.79 \\
GRB 141011282 & 484.6 & 479.28$^{*}$ & 490.72 & 484.48 & 495.66 & 480.09 & 495.34 \\
GRB 150118927 & 474.76 & 467.64 & 475.96 & 470.61 & 482.3 & 466.16$^{*}$ & 481.65 \\
GRB 120811014 & 406.14 & 332.68$^{*}$ & 342.88 & 338.58 & 348.81 & 380.51 & 348.26 \\
GRB 090617208 & 528.9 & 482.82$^{*}$ & 494.72 & 489.34 & 501.63 & 516.92 & 501.41 \\
\hline
\hline
&\multicolumn{7}{c}{Long GRBs} \\  \cline{2-8}
GRB 090424592 & 731.5 & 590.27 & 590.83 & 589.6$^{*}$ & 592.11 & 591.88 & 591.29 \\
GRB 160802259 & 874.02 & 299.84 & 286.36 & 291.11 & 283.24$^{*}$ & 388.32 & 284.19 \\
GRB 160422499 & 674.42 & 647.3 & 588.9 & 625.6 & 592.38 & 575.86$^{*}$ & 596.75 \\
GRB 131028076 & 1241.96 & 418.56 & 427.21 & 410.92$^{*}$ & 421.55 & 498.67 & 418.31 \\
GRB 131014215 & 9618.96 & 1085.59 & 698.91 & 751.05 & 615.14 & 4140.4 & 605.23$^{*}$ \\
GRB 081009140 & 705.99 & 400.86 & 400.32$^{*}$ & 406.76 & 405.21 & 423.72 & 568.6 \\
GRB 140206275 & 674.78 & 555.27 & 557.03 & 527.21$^{*}$ & 538.55 & 531.38 & 538.38 \\
GRB 140329295 & 605.67 & 594.26 & 510.29 & 490.02 & 470.81$^{*}$ & 475.17 & 476.88 \\
GRB 180218635 & 527.76 & 324.26 & 322.86 & 308.71 & 307.7 & 322.23 & 302.71$^{*}$ \\
GRB 110921912 & 463.93 & 449.24 & 437.97 & 439.69 & 437.13 & 428.84$^{*}$ & 437.77 \\
GRB 130606497 & 692.45 & 712.17 & 666.94 & 632.2 & 631.58 & 656.47 & 616.49$^{*}$ \\
GRB 150201574 & 973.97 & 705.73 & 665.83 & 642.42$^{*}$ & 654.79 & 654.8 & 700.8 \\
GRB 150118409 & 661.54 & 530.08$^{*}$ & 540.87 & 536.25 & 547.04 & 537.92 & 551.03 \\
GRB 090618353 & 318.0$^{*}$ & 335.95 & 326.0 & 336.35 & 330.72 & 318.35 & 331.09 \\
GRB 171010792 & 468.69 & 492.1 & 458.17 & 458.18 & 437.43 & 424.27$^{*}$ & 428.96 \\
GRB 180113418 & 708.69 & 707.96 & 648.21 & 626.3 & 594.54$^{*}$ & 624.1 & 603.89 \\
GRB 150902733 & 1152.88 & 559.62 & 526.73 & 531.24 & 509.28$^{*}$ & 610.55 & 518.25 \\
GRB 180113011 & 677.4 & 524.13$^{*}$ & 532.24 & 529.99 & 537.92 & 525.92 & 540.14 \\
GRB 100701490 & 493.77 & 413.46 & 415.08 & 395.24$^{*}$ & 403.74 & 403.78 & 405.86 \\
GRB 170409112 & 1157.08 & 634.07 & 592.73 & 619.0 & 592.33$^{*}$ & 611.1 & 600.44 \\
GRB 120426090 & 880.13 & 453.93 & 436.61$^{*}$ & 457.04 & 441.89 & 454.14 & 438.1 \\
GRB 140508128 & 391.64 & 325.25 & 327.41 & 287.21$^{*}$ & 297.07 & 299.83 & 295.34 \\
GRB 150330828 & 581.82 & 603.66 & 592.92 & 571.91$^{*}$ & 583.66 & 578.2 & 579.5 \\
GRB 100719989 & 660.88 & 475.24 & 462.51 & 460.84 & 459.19 & 465.95 & 455.99$^{*}$ \\
GRB 131229277 & 983.59 & 484.71$^{*}$ & 496.63 & 486.5 & 498.7 & 655.4 & 495.23 \\
GRB 180305393 & 781.32 & 578.83 & 567.97$^{*}$ & 579.81 & 576.1 & 594.92 & 584.07 \\
GRB 170115743 & 896.02 & 668.38$^{*}$ & 678.22 & 674.58 & 1119.1 & 693.15 & 683.0 \\
GRB 170527480 & 759.86 & 531.84 & 522.87$^{*}$ & 538.27 & 529.13 & 544.0 & 526.8 \\
GRB 110825102 & 633.42 & 529.06$^{*}$ & 532.76 & 535.24 & 535.8 & 533.92 & 532.0 \\
GRB 130502327 & 1264.4 & 572.84$^{*}$ & 584.27 & 576.73 & 586.78 & 682.54 & 583.45 \\
GRB 101014175 & 370.37$^{*}$ & 442.97 & 405.19 & 401.23 & 385.78 & 372.24 & 389.59 \\
GRB 150510139 & 772.32 & 505.92$^{*}$ & 506.7 & 512.07 & 512.9 & 506.57 & 514.07 \\
GRB 161218356 & 959.21 & 566.44 & 559.43 & 560.0 & 557.42$^{*}$ & 567.05 & 577.48 \\
GRB 120129580 & 833.77 & 397.36 & 338.73 & 351.32 & 328.74$^{*}$ & 359.51 & 336.76 \\
GRB 110625881 & 770.2 & 645.13 & 560.86 & 571.06 & 561.42 & 550.44$^{*}$ & 582.45 \\
GRB 100826957 & 443.94 & 466.43 & 446.02 & 435.02 & 432.24 & 424.87$^{*}$ & 427.15 \\
GRB 130305486 & 761.15 & 536.88 & 534.46 & 529.88$^{*}$ & 537.22 & 561.81 & 537.21 \\
GRB 170808936 & 741.55 & 732.2 & 678.21 & 680.3 & 664.47 & 638.2$^{*}$ & 684.86 \\
GRB 160625945 & 1909.5 & 2699.83 & 1473.65 & 1256.17 & 1038.53 & 1331.1 & 824.75$^{*}$ \\
GRB 090926181 & 1225.21 & 602.96 & 575.16 & 590.93 & 574.46$^{*}$ & 611.98 & 579.78 \\
GRB 140416060 & 286.58 & 281.57$^{*}$ & 292.12 & 287.07 & 296.17 & 286.43 & 301.18 \\
GRB 110301214 & 767.22 & 708.29 & 680.4 & 674.25 & 660.69 & 657.64$^{*}$ & 664.03 \\
GRB 151231443 & 380.89 & 379.98 & 380.22 & 379.26 & 376.28 & 365.65$^{*}$ & 377.56 \\
GRB 130518580 & 597.18 & 640.17 & 599.56 & 600.77 & 602.83 & 576.77$^{*}$ & 602.13 \\
GRB 091003191 & 455.9 & 405.36 & 417.14 & 397.04 & 403.89 & 390.66$^{*}$ & 402.89 \\
GRB 150403913 & 458.33 & 478.33 & 478.51 & 442.39 & 443.45 & 431.85$^{*}$ & 439.45 \\
GRB 090820027 & 1079.08 & 539.73 & 493.36$^{*}$ & 507.85 & 497.72 & 502.53 & 501.37 \\
GRB 090902462 & 2585.62 & 1548.01 & 706.96 & 1554.2 & 712.99 & 3098.02 & 548.12$^{*}$ \\
GRB 171119992 & 290.5 & 287.47 & 297.42 & 290.51 & 299.02 & 284.21$^{*}$ & 298.19 \\
GRB 160910722 & 1816.26 & 729.31 & 650.27 & 557.52 & 548.95$^{*}$ & 880.89 & 561.04 \\
GRB 120711115 & 394.08 & 415.27 & 425.09 & 393.79$^{*}$ & 403.29 & 405.85 & 404.28 \\
GRB 100829876 & 373.09 & 307.73 & 282.73$^{*}$ & 292.48 & 288.57 & 294.39 & 292.43 \\
GRB 150213001 & 866.19 & 783.68 & 756.91 & 732.69 & 725.53$^{*}$ & 748.31 & 886.65 \\
GRB 101123952 & 485.33 & 437.88$^{*}$ & 447.96 & 443.78 & 453.85 & 442.85 & 451.86 \\
GRB 081215784 & 906.74 & 751.13 & 704.97 & 566.27$^{*}$ & 575.45 & 630.0 & 566.94 \\
GRB 131231198 & 716.52 & 609.54 & 622.07 & 596.42$^{*}$ & 607.78 & 615.48 & 603.89 \\
GRB 160720767 & 521.93 & 459.06$^{*}$ & 463.72 & 463.33 & 469.32 & 462.4 & 471.41 \\
GRB 150627183 & 510.46 & 479.38 & 469.48 & 460.21 & 468.68 & 451.13$^{*}$ & 473.14 \\
GRB 160530667 & 1341.14 & 732.46 & 664.92 & 652.24 & 650.94 & 663.68 & 642.24$^{*}$ \\
GRB 140102887 & 819.08 & 554.27 & 548.88 & 527.02$^{*}$ & 536.78 & 536.6 & 718.96 \\
GRB 160509374 & 674.09 & 666.91 & 659.04 & 604.77$^{*}$ & 616.18 & 613.71 & 615.55 \\
GRB 160821857 & 724.48 & 905.87 & 803.15 & 620.44 & 596.31$^{*}$ & 718.81 & 610.33 \\
GRB 150210935 & 849.99 & 595.01 & 576.19$^{*}$ & 599.31 & 580.34 & 582.13 & 586.18 \\
\enddata

\end{deluxetable}

\bibliographystyle{aasjournal}
\bibliography{references}

\end{document}